# Conceptualizing and Realizing A Smart City Model for Bangladesh

Nafees Mansoor

University of Liberal Arts Bangladesh (ULAB), Dhaka, Bangladesh
nafees@ieee.org

**Abstract.** The outbreak of the novel Corona Virus in 2019, named COVID-19, causes the ongoing global pandemic. This pandemic has a devastating socio-economic impact across the globe. On the other hand, due to the pre-pandemic aggressive urbanization with the steep population growth, modern cities also facing substantial challenges. Hence, usage of technology is anticipated to be the precondition for adaptive, resilient, and sustainable development, where, a smart city is defined as the accumulated advanced ideas of information and technology aiming to ensure a decent quality of life to the inhabitants. Considering the current growth rate of 1.9 percent, the projected population of Bangladesh will exceed 180 million in 2026. It is also speculated that Dhaka being the capital will be populated with 14 million inhabitants. Moreover, Dhaka has already been labeled as the most densely populated city in the world. Thus, concerned authorities are facing enormous challenges to provide and ensure fundamental services to the inhabitants. Therefore, this has become the need of the hour to conceptualize an information and communication technology-driven smart city for Bangladesh. A smart city may contain numerous components; however, the proposed framework identifies seven components and services for smart cities in Bangladesh. These are healthcare, education, transportation, public safety, real estate, utilities, and city administration. Discussions on these components are carried out in this paper.

**Keywords:** Smart City, Post COVID-19, Sustainable Wellbeing, New Normal, 2020 Pandemic

## 1 Introduction

Present-day cities are defined as the melting pot for individuals of different backgrounds. On the other hand, with the emergence of the industrial revolution, the geographical boundary and financial importance of modern cities have been rapidly mounting. Thus, with rapid urbanization and sharp growth of population, modern cities are confronted with substantial challenges. Hence, usage of technology is the precondition for adaptive, resilient, and sustainable development. The term "Smart City" has been getting a lot of attention for the past few years. Lately, many cities, mostly in developed countries, have initiated, implemented, and transformed into smart cities. Moreover, other nations also have started to materialize the concept of a smart city.

A smart city is a generic term used to specify a city's ability to respond as promptly as possible to the needs of the citizens and to create a knowledge infrastructure. Thus, a smart city can be defined as a place where modern technologies are translated to provide better public services for citizens while ensuring efficient usage of resources with minimum environmental impact. A smart city's core architecture is divided into eight major clusters - organization, technology, governance, policy, people and communities, economy, built infrastructure, and the natural environment [1]. Moreover, many aspects of day-to-day life can be added to the definition of a smart city such as sustainable and renewable energy, waste management, law enforcement, and security, efficient power plants, smart architecture, economical infrastructures, etc.

Though, the term smart city is sometimes used to define "intelligent city" or "digital city" or "virtual city", there are some distinct differences among these concepts. An intelligent city can be defined as a city that practices and implements a higher level of knowledge and innovation [2]. On the other hand, digital cities are cities, which are not made of steel and concrete, but of computers, wireless links, electronic connections, and bits. In this sense, without any physical dimensions, a city can be digital resembling "real" cities in various ways. Virtual cities in contrast are the cities produced in games and other forms of electronic media.

On the other hand, a smart city is defined as the accumulated advanced ideas of information and technology aiming to ensure a decent quality of life for the inhabitants. Thus, in such cities, information technology is combined with infrastructure, architecture, and everyday objects to solve social, economic, and environmental challenges. Therefore, a city can be defined as "smart" when human intelligence, social structures, traditional communication infrastructure, and modern communication technology are fused for sustainable economic development and uplifting quality of life [3]. In other words, a smart sustainable city is an innovative city that uses information and communication technologies (ICTs) to improve living standards by ensuring efficient operations and services. Such cities also meet the needs of present and future generations with respect to economic, social, and environmental aspects. Therefore, the key points of a smart city can be divided into four categories, namely sustainability, quality of life, urban aspects, and intelligence or smartness [4].

Modern cities are encountering substantial challenges due to escalated population growth and rapid urbanization of rural areas. These challenges can be addressed effectively and efficiently through smart city technology. Thus, implementing information and the technology-driven smart city can improve the transportation system. Moreover, smart cities are considered to ensure the safety and security of citizens, particularly women, children, and the elderly. Another technology that will make smart cities more efficient is the Internet of Things (IoT). IoT refers to appliances and devices connected to the Internet. Numerous household appliances such as refrigerators, locks, washing machines, lights, and HVAC (Heating, Ventilation, and Air Conditioning) units can be programmed to become IoT compliant. These appliances can be used to observe and provide surrounding feedback or do specific tasks. Besides, technology can be useful to improve efficiency in urban transport systems which are prone to bottlenecks traffic jams,. GPS systems, cameras, and traffic light coordination systems can be monitored, and traffic light can be controlled according to the feedback.

The rest of this paper is organized as follows. An overview of the smart city in a global context is presented in Section II. Section III discusses the importance of implementing smart city concepts in Bangladesh. Later, in Section IV, the proposed framework is presented and the paper is concluded in Section V.

## 2 Smart City in Global Context

Intending to create a smart city, some of the countries have already implemented the concept of a smart city. Many cities in South Korea, Sweden, UK, the USA, Japan, Canada, and Singapore have topped the smartest city list [5]. At the same time, numerous other cities implementing the smart city strategy. For instance, Helsinki is trying to emerge as a smart city cluster, where Helsinki is particularly prioritizing the usage of mobile and wireless communications [6]. Lisbon, a European smart city, is focusing on improving the city's entertainment and quality of living by actively involving citizens in the city's governance model [7]. Lisbon plans to become an international hub by bridging surrounding American, European and African cities. Manchester, a fast-growing city in the UK, is promoting community engagement and social capital by adopting modern technologies [8]. New Songdo, an international business city in South Korea, has set its first objective to use ubiquitous computing in the city while Osaka, a commercial seaport city in Japan, is implementing ubiquitous information systems in the city [9, 10]. In Europe, Barcelona has the view to implement ICT to pursue social and urban growth, where the smart city concept is used as a strategic tool [11]. Another European city, Oulu in Finland, is becoming an innovating technological city to lead other smart cities in and Northern Europe [12].

In India, the government lead by Prime Minister Narendra Modi sets the vision to build 100 new smart cities in order to spur economic growth and urbanization [13]. Thus, the Indian Prime Minister in a speech quoted, "Cities in the past were built on riverbanks. They are now built along highways. But in the future, they will be built based on the availability of optical fiber networks and next-generation infrastructure." [14]. INR 70.6 billion (US$1.2 billion) has been allocated by the Indian Government to developing smart cities in the 2014–15 budget, where the strategic components of area-based development are city improvement, renewal, and extension. India also plans to add a Pan-city initiative in which smart solutions can be applied to cover larger parts of the city.

## 3 Importance of Smart City Adaptation in Bangladesh

More than 7 billion people reside on earth. In 1950, the population was only about 2.5 billion, stipulating that the population growth rate has been 180% in the last 60 or so years. Considering the current growth rate of 1.9 percent, it is projected that the total population of Bangladesh will remain in the range of 170 to 180 million in 2021 (Table 1). With this rate, it is also projected that by then the population of Dhaka, the capital city, will be close to 14 million and the population of Chittagong, the largest coastal seaport city, will be around 9 million [15].

**Table 1.** Projected total population (in thousands) [15]

| Year | National | Chittagong | Dhaka |
|---|---|---|---|
| 2011 | 149764 | 7914 | 12516 |
| 2016 | 160221 | 8440 | 13142 |
| 2021 | 171684 | 8990 | 13798 |
| 2026 | 182096 | 9460 | 14366 |
| 2031 | 190735 | 9805 | 14777 |
| 2036 | 196299 | 10080 | 15081 |
| 2041 | 201314 | 10295 | 15291 |
| 2046 | 205255 | 10429 | 15372 |
| 2051 | 207869 | 10474 | 15323 |
| 2056 | 209466 | 10456 | 15193 |
| 2061 | 209415 | 10343 | 14936 |

Moreover, the population density growing rate is another key concern for Bangladesh. Bangladesh has been considered as one of the top highly dense populated countries where Dhaka has been branded as the most densely populated city in the world and Chittagong ranks 6th in the list. Currently, the population density of Dhaka is approximately 112,700 per square km and for Chittagong, the number is 73,900 per square km. Thus, the Government of the country is facing enormous challenges to provide and to ensure the very fundamental services to its inhabitants.

Furthermore, this rising population also posing negative impacts on areas like climate change, food supplies, infrastructure, and much more. Water crises such as water scarcity and water quality have become critical in urban areas. While Bangladesh has made significant progress in providing safe water to its citizens, the gross disparity in coverage still persists. Hence, waterborne diseases have become acute in the country, where only diarrheal diseases are killing over 100,000 children each year [16]. The following parts of the paper discuss a few major problems in Bangladesh, which is somewhat triggered by population density.

Being prone to flooding, Bangladesh has been repeatedly exposed to severe flooding. For instance, 21 percent of the country gets engulfed by floods water annually [17]. Moreover, the floods of 1970, 1974, 1980, 1987, 1988, 1996, 1998, 2004, 2007, and 2009 had significant impacts on the countries economy. During the flood of 1988, 60 percent of Bangladesh was submerged under water affecting 70 million people, whereas, in 1998, 68% of the country was again flooded [17]. During floods, the water level stays up to 60 cm on a lot of roads. Floods in Dhaka are either caused by heavy rainfall in the city or by the overflow of the surrounding rivers. The western part of Dhaka is safeguarded from river flooding by elevated streets and by an embankment built after the devastating 1988 flood. Most of the city expansion is taking place in the eastern part of Dhaka consists of low-lying floodplains that submerge in water during the monsoon season. Meanwhile, a coastal city like Chittagong experience flood almost every year. This is because, when heavy rainfall coincides with a high-water level in the river/sea, water cannot be naturally evacuated through the existing drainage system. Though retention areas are supposed to mitigate such situation while the pumping stations have been set up to evacuate flood water, however, lack of maintenance and lack of coordination among responsible organizations hampered efforts for flood management. Furthermore, the construction of embankments through low-lying areas without providing adequate drainage facilities had caused internal floods [18 - 19].

Additionally, air pollution has become a serious environmental and health hazard. In Bangladesh, air pollution is caused by increasing motorization. Similarly, biomass fuels used for cooking in the poorly ventilated area cause indoor air pollution. While industrial and automobile emissions are the main sources of air pollution outdoor.

The two major industrial areas of Bangladesh are Dhaka and Chittagong, where water pollution is acute. Paper, fertilizer, pesticides, textile, tannery, and pharmaceutical factories are the main contributors to water pollution for these zones. Moreover, a large quantity of untreated industrial wastes is being disposed of in over 200 rivers in Bangladesh. Furthermore, about 700 tanneries in Dhaka city are disposing approximately of 16,000 cubic meters of toxic wastes [20].

The noise pollution is another major crisis in Bangladesh. World Health Organization (WHO) proposes that 60 decibels (DB) sound can cause temporary deafness and 100 DB sound can make a person completely deaf. According to the Noise Pollution Control Rules 2006, in less dense areas, during the day, the ideal sound condition for Bangladesh is 50 dB and during the night, 40 dB. In residential areas, at daytime 50 dB would be ideal and at night 45 dB [21]. Industries, construction works, motorized vehicles, and indiscriminate use of loudspeakers are the sole contributors to noise pollution. At present, the noise level in Dhaka city is estimated ranging from 80 to 120 decibel that has already impaired 10 percent of the city commuters [21]. If the present situation continues, major populations in the urban areas of Bangladesh will lose 30 decibels of hearing power.

Thus, scarcity of resources, inadequate and deteriorating infrastructure, health concerns, and demand for better economic opportunities are the main concerns for Bangladeshi cities.

## 4   The Proposed Framework

With the advent of the industrial revolution, business globalization and financial activities, the importance of modern cities has been rapidly rising. Thus, with the need for advancing urbanization and accelerated population growth, modern cities are facing substantial challenges. In a smart city, government, businesses, and communities are required to rely on technology to overcome the challenges of rapid urbanization. Thus, a smart city is a combination of software systems, state-of-the-art technologies, network infrastructure, and client devices. To successfully deliver on the smart city vision, the key drivers of such a city need to be identified initially. Moreover, establishing a connected city with its components to collect data is the initial step for a smart city, the proposed framework associates seven components and services for a smart city in Bangladesh. These components are healthcare, education, transportation, public safety, real estate, utilities, and city administration (Figure 1). A discussion on these components is carried out in the following subsections.

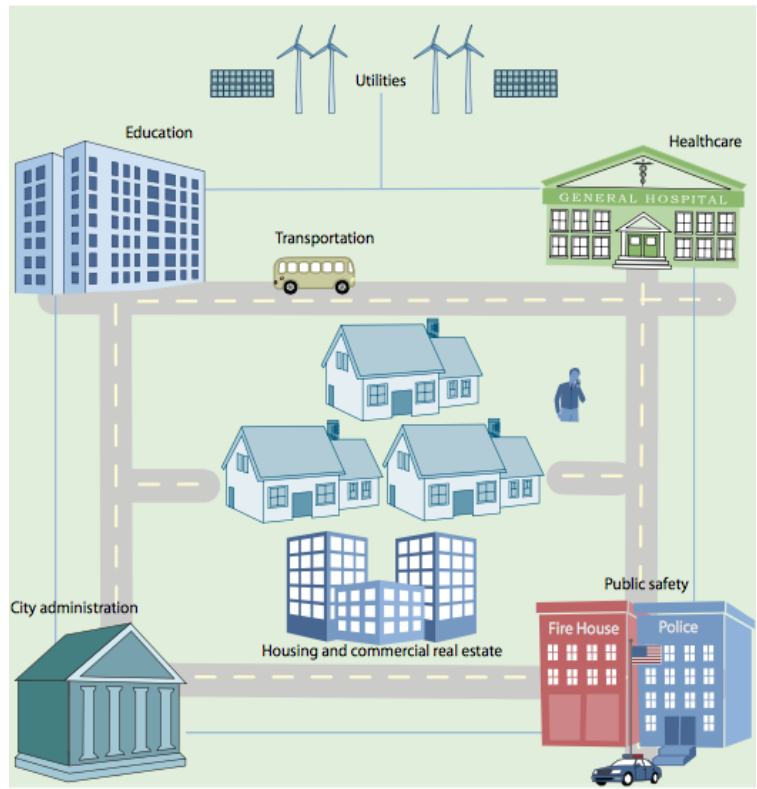

**Fig. 1.** Components of the proposed framework [22].

### 4.1 Smart Healthcare

Now a day, advances in information and communication technologies have enabled better lifestyles for the inhabitants of cities. Moreover, with the rapid growth of the Internet of Things (IoT), healthcare services are also getting smarter and efficient. Therefore, smart cities have an important role to play in the advancement of healthcare management. In such cities, advanced sensor infrastructures can provide continuously updated information about pollution, noise, temperature, etc. This live data provides the foundation for Smart Cities to develop their health applications that provide alerts and advice for citizens [23]. Moreover, such healthcare facilities can also collect data and can use it for population health management. After analyzing the data, a smart healthcare system can identify potential localized outbreaks of illness and epidemics. For example, when existing air quality conditions have the potential to negatively impact known health issues, an app can suggest route alternatives so citizens can avoid specific areas of a city. Furthermore, a smart healthcare system can be built on scalable storage systems and a communications platform. With this type of IT foundation, patient records are electronically stored and shared wherever they are needed. The communication platform enables quick response to emergency services. Video conferencing technologies facilitate remote medical center services to patients' homes, for those who cannot travel to hospitals. For instance, elderly and chronically ill individuals can be monitored and nursed efficiently using different sensors and connected devices. Hence, smart cities will take medical care to the next level enabling connectivity across devices and remote monitoring of patients.

With the predicted increase of urban residency, cities in the country need to expand healthcare infrastructures and need to improve the interoperability of networks. Thus, one of the key elements of the proposed smart cities in Bangladesh is to implement smart healthcare facilities. Moreover, to enhance infrastructure facilities, the proposed smart city may set up a "Healthcare Network" across the city. This network can bridge the gap in the demand and supply of beds in the city and can provide access to good quality medical care to the citizens.

Instead, contemporary wearable devices and sensors can measure heart rates, quantity and quality of sleep, steps covered, and workout intensity. Moreover, sophisticated monitoring devices can observe and report blood oxygenation level, hydration, lung capacity, sugar levels in the blood, Body Mass Index (BMI), and capillary blood flow. Furthermore, the capability to dispense, adjust, and control medication dosages is also offered in smart pill bottles. Therefore, a mobile app can be introduced in the proposed smart cities where the proposed health-monitoring app will ensure that the physicians to get updated information about the health conditions of the patients remotely. This app can also offer users the convenience of getting alerts for medication and doctor's appointments. Users can also set up notifications and workflows based on health status which will proactively take actions. For instance, in a healthcare facility, electronic medical records can be used to automatically send instructions to its staff to provide medicine to a patient or to take a patient for diagnosis. Similarly, patients receive text messages on their phone to remind them of scheduled appointments.

Access to an advanced healthcare system allows many more people to follow their daily routine and activities, and even manage conditions from home rather than a hospital, improving their overall quality of life. For healthcare providers, to monitor patients and adjust medications and treatments remotely improves the level of healthcare that can be delivered to a greater number of patients. Thus, the potential applications of technology to improve efficiency, safety, and quality of health care in Bangladesh are truly enormous.

### 4.2  Smart Education

Focusing on education, a smart city is termed as the center of making better-educated individuals and skilled workforces. Smart cities act as magnets for creative people and workers, and this allows the creation of a virtuous circle making them smarter and smarter. Consequently, a smart city has multiple opportunities to exploit its human potential [24]. Moreover, in a smart city, heightened use of technology in education aims to increase access. A technology driven education system can also improve the quality of education, while reducing the costs of education. For example, the introduction of PCs and the Internet will increase access to educational resources for all students of every sphere of the society. Likewise, the use of digital content and collaboration technologies can improve the quality of education, at more convenience, at a lower cost. For instance, educational institutes in the proposed smart cities can enhance the accessibility and can reduce the cost of education by using e-learning and m-learning systems.

### 4.3  Smart Transportation

Rapid transition to a highly urbanized population creates challenges in transportation. Therefore, proposed smart cities in Bangladesh require to consider a smart transportation system. The proposed smart transportation system is divided into three sectors. Descriptions of these sectors are provided in the following subsections.

#### a.  *Traffic Congestion Monitoring*

The city traffic management system is the key concern to planners and users. From the end-user viewpoint, congestion is the main parameter to evaluate any traffic management system. It is considered that congestion occurs as the demand for transport facility exceeds the ability to supply the acceptable level of service. From an applied viewpoint, the congested traffic conditions are simply defined as the blockage of roads with queued vehicles. Moreover, congestion can be recurrent or nonrecurring and can be positioned at isolated points or across the network. Recurrent congestion falls under predictable variations in demand associated with morning and evening peaks and other events. On the other hand, non-recurring congestion is fundamentally unpredictable and typically linked with instances like road works, accidents, breakdowns, etc. In such a network, an area-wide phenomenon is a congestion which is considered as the unstable flow condition. From the user point of view, road congestion initiates lower speeds and longer waiting intervals. Since traveling time can be a meaningful and important indicators of congestion to analyze the performance of traffic.

The existing traffic management system in Bangladesh is extremely troublesome where the country is losing millions of work hours due to traffic jams. Besides, people from all walks of life are suffering tremendously. Additionally, air and sound pollutions are considered as the side effects of traffic congestion. Therefore, in the proposed smart city, a smart trafficking system is identified where this system will predict traffic in various periods of time. Moreover, using the cognitive radio enabled vehicular ad-hoc network technology, the system will establish a network of all the vehicles on the road and an artificial intelligence-enabled central system will make traffic decisions [25].

### b. Smart parking

The demand for parking infrastructures is observed to be increasing over the years [26]. In most cities, spotting a vacant parking space at busy hours is becoming more difficult and in cases, it is becoming impossible. Identifying the necessity to resolve the concern and as well as meet the demand for parking spaces with improved services, parking controlling bodies are motivated to implement smart and sustainable solutions. Since the parking availability system operates in real-time, the prospect for the smart parking solutions is anticipated to be integrated with features such as parking spot reservation and dynamic pricing systems. In such a system, smart meters, smart cards, telecommunications, and e-parking may also be incorporated. Hence a smart parking system enables the existing system to satisfy the customers' needs effectively.

Since smart parking detects vehicles, thus the parking garages must be equipped with such devices at the entrances and exits. This will ensure an accurate vehicle count. However, the main challenges for such a system are in installation and maintenance processes. In a smart parking system, video cameras, ultrasonic sensors, and Doppler radars are also implemented for improved effectiveness of the system. Such systems were first implemented in countries like the United Kingdom and Japan, however other countries have implemented such smart parking facilities in most of their major cities [27].

In the proposed smart cities, smart parking technology benefits the customer and the parking operator in various ways. Using this system, a user can determine space availability before entering the garage and/or parking level. The user also can plan for transit to public transportation with such smart parking systems. Moreover, the parking operator can use the system data to develop or improve pricing strategies and to predict future parking patterns and trends. The parking operator can use this system data to prevent vehicle thefts. The system significantly aims to reduce traffic and the resulting vehicle emissions– by decreasing the time required for users to locate open spaces.

### c. Smart Street Lighting System

The existing street lighting system wastes energy in a large scale. However, increasing availability of flexible lighting technology such as Light Emitting Diode (LED) and wireless communication technologies, effective street lighting systems are becoming more realistic.

Thus, the proposed smart cities will provide a Smart Street Lighting (SSL) system. The SSL system will dynamically switch street lamps based on pedestrians' locations and desired safety. In the proposed system, each pedestrian is localized via the Smartphone which sends the location and configuration information periodically to a SSL server. In a SSL, all the street light posts are equipped with ZigBee based device to receive instruction from a SSL server via multi-hop routing technique.

### 4.4 Public Safety and Public Health

With the increasing number of inhabitants, public safety ensuring agencies such as police, firefighters, etc. require more rapid responses during emergency situations and need to be agile all the time to control crime. At present, a smart public safety agency operates on real-time information. Broadly, todays' cities are designed to establish a network for public safety with earliest dispatch and video analytics. Such system will also have the ability to simulate an anticipated event with geo-localization information. On the other hand, water pollution monitoring, noise urban mapping, temperature monitoring, waste management are some other important areas for public health in the proposed smart city. The water pollution monitoring system will be designed to measure, monitor and analyze pollution, pollutants and other factors regarding mostly of open water. Noise pollution is at extreme in Bangladesh. It is creating severe health issues such as raised blood pressure, hearing loss, decreased cognitive functions, annoyance and even psychological symptoms. Hence, the proposed noise monitoring system will measure noise strength and its impact area. More importantly, the system will detect the source of the noise. Thus, the proposed public safety and public health system will use real-time information for prompt response during emergency situations and security threats.

### 4.5 Real Estate

Structural Health Monitoring (SHM) is defined as the damage identification system of an infrastructure. SHM monitors a structure or a mechanical system over a period of time, extracts the damage-sensitive features and statistically analyzes these features to predict the future health of a system. Moreover, a long-term SHM periodically updates the information regarding a structure considering the inevitable aging and damage caused by the functioning surroundings. Such SHM can be used for rapid condition screening [28].
Bangladesh has a long history of structural health damage issues. As a country with rich history, Bangladesh has several historical structures and these are being damaged by various factors. To preserve the heritages and history, necessary steps for detecting damages and preventing from further erosion, SHM should be implemented. In recent years, several buildings have collapsed and many buildings are in danger of being collapsed. Several hundreds of lives were lost and many suffer from the traumas. In 2003, a staggering 1135 people were killed in 5 garments factories collapse [29]. This is due to poor maintenances of the buildings and bad ingredients for the building process. On the other hand, several aircrafts accidents have been reported, where a major cause of these accidents is identified as poor prediction and maintenance system.

Thus, the proposed smart cities may provide a system, which monitors the structural health of buildings (high-rises, hospitals, towers, arenas/stadiums, and historical structures), utility systems (communication towers, power plants, dams, pipelines, water plants), transportation systems (bridges, airports, ports, rail-tracks), etc. This SHM needs to consider thorough and state-of-the-art approaches for damage detection and assessment. Using different types of sensors (acceleration, velocity, displacement, tilt, strain, force, pressure, wind, GPS, temperature, humidity, corrosion, etc.), this scheme will be equipped with automated fault detection and real-time monitoring features.

### 4.6 Utilities

It has been already identified that the upcoming energy demand requires innovative grid management. Hence, the concept of Smart Grid (SG) is defined as a smart system that uses ICT and power electronics for efficient energy distribution. SG also balances the energy production and energy consumption autonomously. However, SG is also defined based on the operating features over the technology it uses. Since deployment of SG requires time where the new features will be added on the existing system. Hence, it is anticipated that a SG system will have three major components namely, distributed intelligence, information and communication technologies and an autonomous controlling system. Integrating these three components, a SG system is also assumed to be operated over the renewable energy systems and able to distribute energy more effectively. Again, from a different view point, SG systems are also termed based on three integrated layers namely, application, power and communication.

Energy has become an important factor for the growth of Bangladesh. However, security risk is one of the primitive design challenges for SG with a centralized generation plant. Moreover, wind and photovoltaic (PV) power generations are getting special attention in the domain of Renewable Energy Sources (RES). Thus, existing energy system incorporated with RES may face additional technical challenges linked with security, reliability and quality. Moreover, the current grid system is also limited with disintegrated architecture, inadequate bandwidth for two-way communications, system components interoperability scarcity and data management incapability. Moreover, almost all of the areas are affected by load shedding due to excessive power consumption, as load balancing is not efficient. Besides, devices malfunctioning, damage of devices or accidents due to improper power delivery is a common phenomenon. Additionally, connection of power station for a distributed power delivery mechanism is absent. Therefore, implementing smart grid in the proposed smart cities is also essential.

### 4.7 City Administration

In the current service-driven economy, a smart city administration needs to be well-informed regarding the city's condition and equipped with sufficient systems to reach the inhabitants effectively. Using Information and Communication Technologies (ICT), a central operating center of a smart city monitors and regulates the health-care and education services. Moreover, this center also coordinates and controls the transportation and utility services. However, to do so, the central city administration needs to establish coordination among all its service providing agencies. It is worthwhile to mention that some of these agencies already have sufficient precise data, however due to lack of coordination these data are less utilized or underutilized. It is also observed that since there is no coordination among the agencies, they are collecting and maintaining redundant data. Hence, a smart administration may collect data from one agency and supply these data to other agencies. However, in order to ensure an effective smart city administration, common view points among these agencies needs to be introduced [30].

Similar to FixMyStreet.com [31], proposed smart cities of Bangladesh may introduce an app which allows the inhabitants to report local issues such as fly tipping, graffiti, cracked slabs or damaged street lights, and to track their resolution. Moreover, this app may also enable the citizens to report about any crime and any criminal/ terrorist, where an automatic alert is delivered to the concerned agencies.

## 5    Conclusion

In modern cities, consequential challenges have been introduced by the process of aggressive urbanization and swelling growth in population. Hence, usage of technology is anticipated to be the precondition for adaptive, resilient, and sustainable development. Therefore, the smart city in the proposal is defined as the accumulated advanced ideas of information and technology aiming to ensure a decent quality of life for the inhabitants. The proposed framework establishes seven components and services for a smart city in Bangladesh. These components are healthcare, education, transportation, public safety, real estate, utilities, and city administration. It is also anticipated that the proposed framework connects these components and collects data from the mentioned components. Since establishing a connected city is the initial step for a smart city, hence our future research will focus on managing data and building the platform to connect these data with different applications. Thus, with proper guidance from the academics and the researchers, and corporation from the governing administration and the people, smart cities can be built in Bangladesh.

**References**

1. Chourabi, H., Nam, T., Walker, S., Gil-Garcia, J. R., Mellouli, S., Nahon, K., & Scholl, H. J. Understanding smart cities: An integrative framework. In System Science (HICSS), 2012 45th Hawaii International Conference on (pp. 2289-2297). IEEE.
2. Komninos, N. (2002). Intelligent cities: innovation, knowledge systems, and digital spaces. Taylor & Francis.
3. Caragliu, A., Del Bo, C., Nijkamp, P.: Smart cities in Europe. Series Research Memoranda 0048. VU University Amsterdam, Faculty of Economics, Business Administration and Econometrics (2009).

4. Nam, T., & Pardo, T. A. (2011, September). Smart city as urban innovation: Focusing on management, policy, and context. In Proceedings of the 5th International Conference on Theory and Practice of Electronic Governance (pp. 185-194). ACM.
5. R.G. Hollands, 2008. "Will the real smart city please stand up? Intelligent, progressive or entrepreneurial?". City, 12(3), pp.303-320.
6. Hielkema, H., & Hongisto, P. (2013). Developing the Helsinki smart city: the role of competitions for open data applications. Journal of the Knowledge Economy, 4(2), 190-204.
7. Wimmer, M. A., Cristiano C., and Xiaofeng M. "Developing an e-government research roadmap: method and example from E-GovRTD2020." International Conference on Electronic Government. Springer, Berlin, Heidelberg, 2007.
8. Williams, K. "Social networks, social capital, and the use of information and communications technology in socially excluded communities: a study of community groups in Manchester, England." (2005).
9. Hyang-Sook, C., Byung-Sun, C. and Woong-Hee, P., Ubiquitous-City Business Strategies: The Case of South Korea. In: Management of Engineering and Technology (PICMET 2007), IEEE (2007).
10. Anthopoulos, L. and Panos E. "From digital to ubiquitous cities: Defining a common architecture for urban development." Intelligent Environments (IE), 2010 Sixth International Conference on. IEEE, 2010.
11. March, H., and Ramon R.F.. "Smart contradictions: The politics of making Barcelona a Self-sufficient city." European Urban and Regional Studies 23.4 (2016): 816-830.
12. Salo, M. "High-tech centre in the periphery: The political, economic and cultural factors behind the emergence and development of the Oulu ICT phenomenon in northern Finland." Acta Borealia31.1 (2014): 83-107.
13. Datta, A. New urban utopias of postcolonial India 'Entrepreneurial urbanization' in Dholera smart city, Gujarat. Dialogues in Human Geography, 5(1), 3-22, 2020.
14. Sen, B. Digital Politics and Culture in Contemporary India: The Making of an Info-Nation (Vol. 16). Routledge, 2016.
15. Bangladesh Bureau of Statistics, Population Projection of Bangladesh for 2016-2061.
16. "Diarrheal Diseases in Bangladesh." Diarrheal Diseases in Bangladesh - Stats, Demographics Affected, Graphiq.
17. AM, Dewan, M. Nishigaki, and M. Komatsu. "Floods in Bangladesh: A comparative hydrological investigation on two catastrophic events." 岡山大学環境理工学部研究報告 8.1 (2003): 53-62.
18. M. Shah, Alam Khan. "Stormwater Flooding in Dhaka City: Causes and Management" (PDF). Institute of Water and Flood Management, Bangladesh University of Engineering and Technology, Dhaka
19. Chusit Apirumanekul; Ole Mark (2001). "Modelling of Urban Flooding in Dhaka City" (PDF). 4 th DHI Software Conference. p. 102.
20. Alam, GM Jahangir. "Environmental pollution of bangladesh–it's effect and control." Pulp and Paper 51 (2009): 13-7.
21. Unb. "Dhaka City: Noise Pollution Turns Acute." The Daily Star, 7 Oct. 2017.
22. Washburn, D., Sindhu, U., Balaouras, S., Dines, R. A., Hayes, N., & Nelson, L. E. (2009). Helping CIOs understand "smart city" initiatives. Growth, 17(2).
23. Boulos, M. N. K., & Al-Shorbaji, N. M. (2014). On the Internet of Things, smart cities and the WHO Healthy Cities. International journal of health geographics, 13(1), 1.
24. [24] Winters, J. V. Why are smart cities growing? Who moves and who stays. Journal of regional science, 51(2), 253-270, 2011.

25. [25] S. M. Nadim Uddin, Nafees Mansoor, Sazzad Hossain,"Cognitive Radio Enabled VANET for Multi-agent Based Intelligent Traffic Management System", 1ST International Conference on Advanced Information and Communication Technology 2016 (ICAICT 2016), May 16-17, 2016, Chittagong, Bangladesh.
26. [26] N. J. Farin, M.N.A.A. Rimon, S. Momen, M. S. Uddin and N. Mansoor, "A framework for dynamic vehicle pooling and ride-sharing system," 2016 International Workshop on Computational Intelligence (IWCI), Dhaka, 2016, pp. 204-208.
27. [27] Chinrungrueng, J, Udomporn S., and Satien T.. "Smart parking: An application of optical wireless sensor network." Applications and the Internet Workshops, 2007. SAINT Workshops 2007. International Symposium on. IEEE, 2007.
28. [28] Farrar, C. R., & Worden, K. (2007). An introduction to structural health monitoring. Philosophical Transactions of the Royal Society of London A: Mathematical, Physical and Engineering Sciences, 365(1851), 303-315.
29. [29] Reuters in Dhaka. "Rana Plaza Collapse: 38 Charged with Murder over Garment Factory Disaster." The Guardian, 18 July 2016.
30. [30] Harrison, C., & Donnelly, I. A. (2011, September). A theory of smart cities. In Proceedings of the 55th Annual Meeting of the ISSS-2011, Hull, UK (Vol. 55, No. 1).
31. [31] King, S. F., & Brown, P. (2007, December). Fix my street or else: using the internet to voice local public service concerns. In Proceedings of the 1st international conference on Theory and practice of electronic governance (pp. 72-80). ACM.